\documentclass[aps,prl,preprint,tightenlines,floatfix,showpacs]{revtex4}
\usepackage{graphics}
\usepackage{bm}
\usepackage{amsmath}
\usepackage{amssymb}
\begin{document}
  \title{Microscopic description of elastic and inelastic proton
    scattering from $^{208}$Pb}
  \author{M. Dupuis}
  \author{S. Karataglidis} 
  \author{E. Bauge}
  \author{J.-P. Delaroche}
  \author{D. Gogny}
  \affiliation{Commissariat \`a l'Energie Atomique, D\'epartement de
    Physique Th\'eorique at Appliqu\'ee, Service de Physique
    Nucl\'eaire, BP 12, 91680 Bruy\`eres-le-Ch\^atel, France}
  \date{\today}
  \begin{abstract}
    Information on the  equation of state (EOS) of  neutron matter may
    be gained  from studies of $^{208}$Pb.  Descriptions of $^{208}$Pb
    require credible models of  structure, taking particular note also
    of  the spectrum.  Such may  be tested  by analyses  of scattering
    data. Herein,  we report on such  analyses using an  RPA model for
    $^{208}$Pb  in a  folding model  of the  scattering.  No \textit{a
    posteriori}  adjustment   of  parameters  are   needed  to  obtain
    excellent  agreement  with data.  From  those  analyses, the  skin
    thickness  of  $^{208}$Pb  is constrained  to  lie  in  the  range
    $0.13-0.17$~fm.
  \end{abstract}
  \maketitle
  
  Interest in large-scale nuclear structure models for $^{208}$Pb, and
  the  specification   of  matter  densities   therefrom,  is  topical
  \cite{Br00,Ka02}.    The    neutron    skin    thickness    ($S    =
  \sqrt{\left\langle  r^2_n \right\rangle} -  \sqrt{\left\langle r^2_p
  \right\rangle}$) of  a heavy nucleus is  related to the  radius of a
  neutron  star \cite{Ho01},  by virtue  of the  equation of  state of
  neutron  matter,   and  understanding  this   relationship  requires
  detailed knowledge  of the  requisite proton and  neutron densities,
  for  which $^{208}$Pb  has  been used  as  the example.  There is  a
  proposal to measure the  skin thickness by parity-violating electron
  scattering from $^{208}$Pb  at the Jefferson Laboratory \cite{Je00}.
  Thus  realistic nuclear structure  models are  required in  order to
  understand  neutron densities  in heavy  nuclei as  well  as neutron
  matter.

  Previous  estimates of the  skin thickness  of $^{208}$Pb,  based on
  analyses of  available electron  and nucleon scattering  data, range
  from 0.1 to 0.3~fm \cite{Ka02,Cl03}. But the skin thickness alone is
  not a sufficient constraint  on the models of structure \cite{Ka02},
  and a proper evaluation of  the neutron density requires analyses of
  scattering data,  for which  proton scattering data  from $^{208}$Pb
  are  particularly suited.  The  dominance of  the isoscalar  $^3S_1$
  component  of  the  nucleon-nucleon ($NN$)  interaction  \cite{Am00}
  ensures  that  proton  scattering  probes the  neutron  density  and
  vice-versa. Elastic scattering  probes the neutron density directly;
  inelastic scattering  from $^{208}$Pb probes  the transitions within
  the neutron surface and  may provide additional information by which
  the skin thickness and neutron EOS may be further constrained. It is
  the purpose  of this  letter to use  the Random  Phase Approximation
  (RPA) to obtain the spectrum  of $^{208}$Pb and to evaluate the wave
  functions obtained  therefrom in  analyses of elastic  and inelastic
  proton scattering  data. Such provides sensitive  constraints to the
  neutron density in $^{208}$Pb.

  Microscopic optical  potentials based  on the Melbourne  $g$ folding
  model  \cite{Am00,La01,Ka02,Kl03}  have  been successfully  used  in
  describing   intermediate  energy  nucleon-nucleus   ($NA$)  elastic
  scattering   without  any   \textit{a   posteriori}  adjustment   of
  parameters. Therein, the optical potential for elastic scattering is
  obtained  from the  folding of  the $NN$  $g$ matrices  for infinite
  matter with  the density  matrix of the  ground state of  the target
  nucleus.  However, the  reliability of  those optical  potentials so
  obtained  rests  upon  the  specification  of a  credible  model  of
  structure for the  target. As the optical potentials  are found from
  the  folding  of effective  $NN$  interactions,  which are  one-body
  operators with respect to the target nucleons, a requirement is that
  the chosen  model of structure exhibits nucleon  degrees of freedom.
  Excellent agreement with  data has been achieved in  the analyses of
  elastic  scattering of  65 and  200~MeV data  across the  mass range
  \cite{Am00} and of exotic  nuclei from hydrogen \cite{Ka97,La01}. In
  the case of $^{208}$Pb,  analyses of elastic scattering data allowed
  discrimination between  disparate models of  $^{208}$Pb which, while
  predicting  the  same  skin  thickness, predicted  different  matter
  densities  which were  reflected  in the  calculated cross  sections
  \cite{Ka02}.

  Excellent agreement  also has been achieved  in describing inelastic
  scattering  in light  nuclei self-consistently  when using  the same
  effective $g$  matrix as the transition operator  within a distorted
  wave approximation (DWA) for the scattering, and when the transition
  density  matrix  elements  are  obtained from  the  same  underlying
  structure models.  This has been  illustrated only for  light nuclei
  \cite{Am00}.   Of   particular   note   is  the   case   of   $^6$He
  \cite{La01,Am00}:  the  neutron  halo  in $^6$He  was  unambiguously
  established  only with  a  self-consistent analysis  of elastic  and
  inelastic scattering data.

  With the  RPA it is  possible to describe inelastic  scattering from
  heavy nuclei with the same level of agreement as that for scattering
  from  light nuclei.  This allows  for  evaluation of  the models  of
  nuclear structure  for heavy nuclei, for which  the specification of
  excited  states and of  transitions to  them is  possible, providing
  additional  constraints to the  neutron density.  Using the  RPA and
  quasi-particle RPA (QRPA) models  for heavy nuclei to obtain density
  matrices  is akin  to  the use  of  no-core shell  models for  light
  nuclei.  Herein, we  shall use  the Melbourne  $g$ folding  model to
  obtain microscopic optical potentials for use in scattering analyses
  and thus evaluate the wave functions obtained from the RPA.

  The  microscopic  $g$-folding  optical  potential for  $NA$  elastic
  scattering  has  been  described  in  detail  in  a  review  article
  \cite{Am00}; we  present a brief  summary of the  model illustrating
  the salient points with regards to the nuclear structure.

  The microscopic,  nonlocal, optical  potential is obtained  from the
  effective $NN$  $g$ matrices in  infinite matter. $NN$  $g$ matrices
  for  infinite matter are  solutions of  the Bruckner-Bethe-Goldstone
  equation  in momentum space  \cite{Am00}, and  are derived  from the
  Bonn-B potential \cite{Ma87}  for the calculations presented herein.
  We obtain effective $NN$ $g$ matrices in coordinate space for finite
  nuclei  whose  Fourier  transforms  best map  those  momentum  space
  (infinite  nuclear matter)  values.  The effective  $g$ matrices  so
  obtained,  which contain  central, tensor,  and  two-body spin-orbit
  terms,  are folded with  the ground  state density  matrix elements,
  obtained from  the assumed model  of structure, to give  the optical
  potential for  elastic scattering. The optical  potential so defined
  is  complex,   energy-dependent  and  nonlocal.  It   has  the  form
  \cite{Am00}
  \begin{multline}
    U(  \mathbf{r},  \mathbf{r}';  E  ) =  \delta\left(  \mathbf{r}  -
    \mathbf{r}' \right) \\
    \sum_{\alpha_1 m_1 \alpha_2  m_2} \rho_{\alpha_1 m_1 \alpha_2 m_2}
    \int  \varphi^{\ast}_{\alpha_1  m_1}(\mathbf{s}) g_D(  \mathbf{r},
    \mathbf{s};  E   )  \varphi_{\alpha_2  m_2}  (   \mathbf{s}  )  \;
    d\mathbf{s} \\
    +  \sum_{\alpha_1 m_1  \alpha_2 m_2}  \rho_{\alpha_1  m_1 \alpha_2
    m_2} \varphi^{\ast}_{\alpha_1 m_1}(  \mathbf{r} ) g_E( \mathbf{r},
    \mathbf{r}'; E ) \varphi_{\alpha_2 m_2}( \mathbf{r}' ) \\
    =  U_D(  \mathbf{r}; E  )  \delta\left(  \mathbf{r} -  \mathbf{r}'
    \right) + U_E( \mathbf{r}, \mathbf{r}'; E )\; ,
  \end{multline}
  where  the  subscripts  $D,E$  designate  the  direct  and  exchange
  contributions,  respectively,  $\alpha  \equiv   \{  n,  l,  j  \}$,
  corresponding to the occupied single-particle orbits, and
  \begin{equation}
    \rho_{\alpha_1  m_1 \alpha_2  m_2} =  \left\langle  \Psi_{J_i M_i}
    \left| a^{\dag}_{\alpha_1 m_1}  a_{\alpha_2 m_2} \right| \Psi_{J_i
    M_i} \right\rangle \, ,
  \end{equation}
  is  the density  matrix.  (For $^{208}$Pb,  $J_i  = M_i  = 0$.)  The
  coordinates   $\mathbf{r}$    and   $\mathbf{r}'$   are   projectile
  coordinates. The  coordinate-space code DWBA98  \cite{Ra98} has been
  used to obtain the results presented herein.

  Inelastic scattering is obtained in  the DWA using the effective $g$
  matrices,  specified  for  elastic  scattering,  as  the  transition
  operators. The DWA transition amplitudes can be written as
  \begin{multline}
    T^{M_fM_i\nu'\nu}_{J_fJ_i}(\theta)          =         \left\langle
    \chi^{(-)}_{\nu'}(\mathbf{k_o}0)        \right|       \left\langle
    \Psi_{J_fM_f}( 1 \cdots  A ) \right| A\mathbf{g}_{\text{eff}}(0,1)
    \\
    \mathcal{A}_{01}  \left\{  \left|\chi^{(+)}_{\nu}( \mathbf{k}_i  0
    )\right\rangle  \left| \Psi_{J_iM_i}( 1  \cdots A  ) \right\rangle
    \right\} \;,
  \end{multline}
  where    the    distorted   wave    functions    are   denoted    by
  $\chi^{(\pm)}_\mu(\mathbf{k}q)$ for an incoming/outgoing proton with
  spin projection  $\mu$, wave vector $\mathbf{k}$  and coordinate set
  `$q$' (either  0 or  1). The nuclear  wave functions are  denoted by
  $\Psi_{JM}(  1 \cdots  A  )$ and,  since  all pairwise  interactions
  between the  projectile and every  target nucleon are assumed  to be
  the  same, it  is convenient  to make  a cofactor  expansion  of the
  nuclear wave  functions from which the  transition amplitudes expand
  to the form, for spin-zero targets,
  \begin{multline}
    T^{M_fM_i\nu'\nu}_{J_fJ_i}(\theta) = \sum_{\alpha_1\alpha_2m_1m_2}
    \frac{(-1)^{j_1 - m_1}}{\sqrt{2J_f + 1}}  \\
    \left. \left\langle j_2\, m_2 \, j_1 \, -\!m_1 \right| J_f \, M_f
    \right\rangle \left\langle J_f \left\| \left[ a^{\dagger}_{j_2}
    \times  \tilde{a}_{j_1} \right]^{J_f} \right\| 0 \right\rangle \\
    \times \left\langle \chi^{(-)}_{\nu'}(\mathbf{k}_o0) \right|
    \left\langle \varphi_{\alpha_2m_2}(1) \right|
    \mathbf{g}_{\text{eff}}(0,1) \\
    \mathcal{A}_{01} \left\{ \left|
    \chi^{(+)}_{\nu}(\mathbf{k}_i 0) \right\rangle \left|
    \varphi_{\alpha_1m_1}(1) \right\rangle \right\} \, .
  \end{multline}
  The one-body  transition density  matrix elements are  obtained from
  the relevant structure model.  Exchange terms enter naturally by the
  action of  the two-body antisymmetrisation operator  $A_{01}$ on the
  bound  nucleon and  projectile  in the  initial  state. The  optical
  potentials  and observables obtained  therefrom are  also calculated
  using DWBA98 \cite{Ra98}.

  The densities for the ground  state of and transitions in $^{208}$Pb
  were obtained from  an RPA calculation using the  D1S effective $NN$
  force  of Gogny  \cite{Bl77,Be90},  which is  density dependent  and
  includes   finite-range  exchange   terms.   That  is   used  in   a
  self-consistent  RPA theory  to obtain  the relevant  density matrix
  elements and  single-particle wave  functions. The RPA  accounts for
  ground state correlations  induced by collective excitations, beyond
  the Hartree-Fock approximation, and  allows for the specification of
  transitions to excited states. For comparison in elastic scattering,
  we  have also  used  densities obtained  from a  Skyrme-Hartree-Fock
  calculation using  the SkM* interaction \cite{Ba82}.  That model was
  deemed most appropriate  for the description of the  ground state of
  $^{208}$Pb based on analyses  of the neutron skin thickness, elastic
  electron scattering  data and elastic proton  and neutron scattering
  data \cite{Ka02}.  The predicted skin thickness from  the SkM* model
  is  0.17~fm  \cite{Ka02} while  that  from  our  RPA calculation  is
  0.13~fm.

\begin{figure}
    \scalebox{0.4}{\includegraphics*{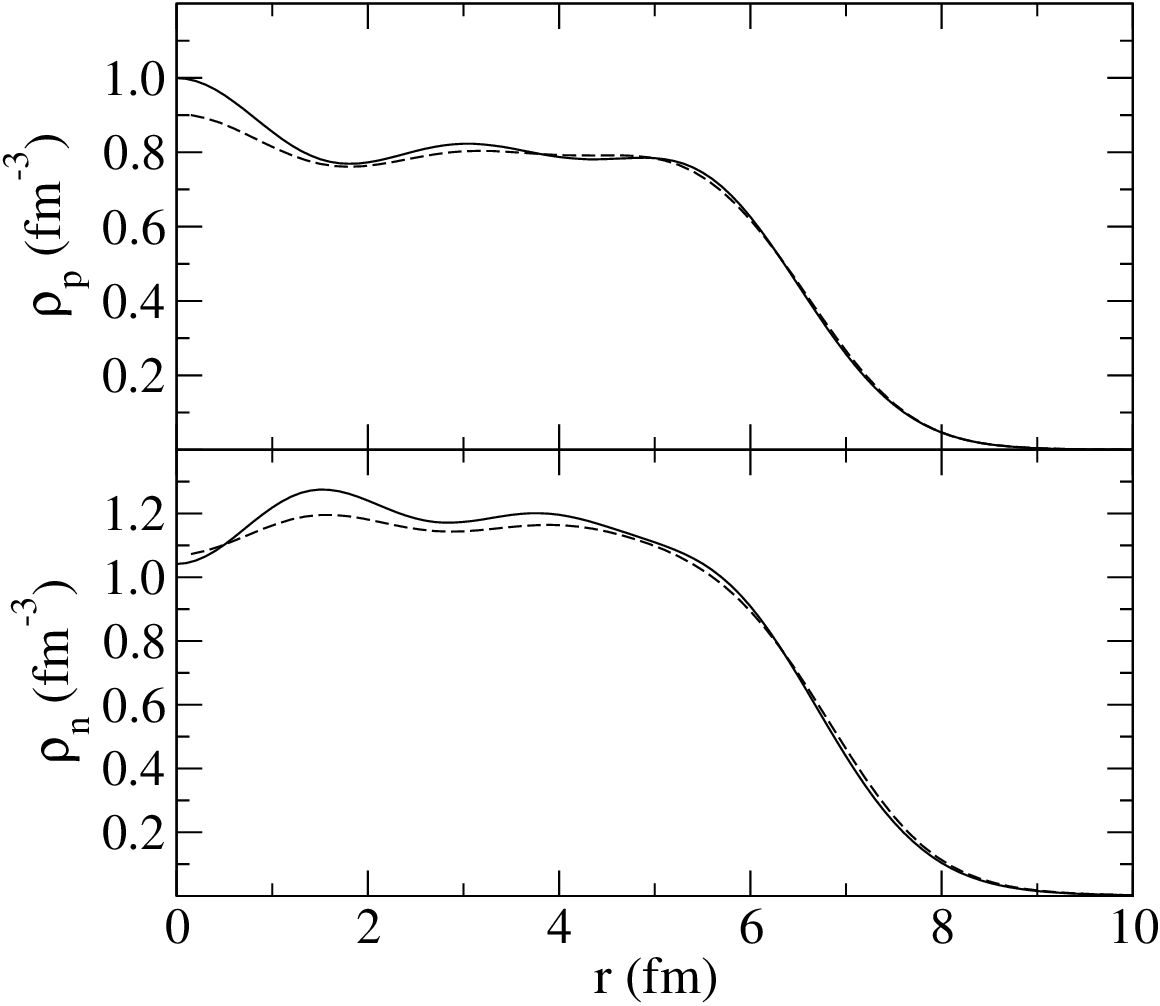}}
    \caption{\label{densities}  Proton   (top)  and  neutron  (bottom)
    densities for  $^{208}$Pb. The results  obtained from the  RPA and
    SkM*   models  are   shown  by   the  solid   and   dashed  lines,
    respectively.}
  \end{figure}

  The  densities, from  the  RPA  and SkM*  models  of structure,  and
  normalised   to   proton   and   neutron  number,   are   shown   in
  Fig.~\ref{densities}.
  The  results obtained  from the  RPA and  SkM* models  for  both the
  proton  and neutron  densities largely  agree with  each  other. The
  predicted proton rms  radii are 5.47~fm and 5.45~fm  for the RPA and
  SkM* models, respectively, while the predicted neutron rms radii are
  5.59~fm  and 5.62~fm, respectively.  This significant  difference in
  the neutron  radii is reflected in  the neutron density  only at the
  surface, where the  RPA model predicts a sharper  surface leading to
  the smaller skin thickness.
  
\begin{figure}
    \scalebox{0.4}{\includegraphics*{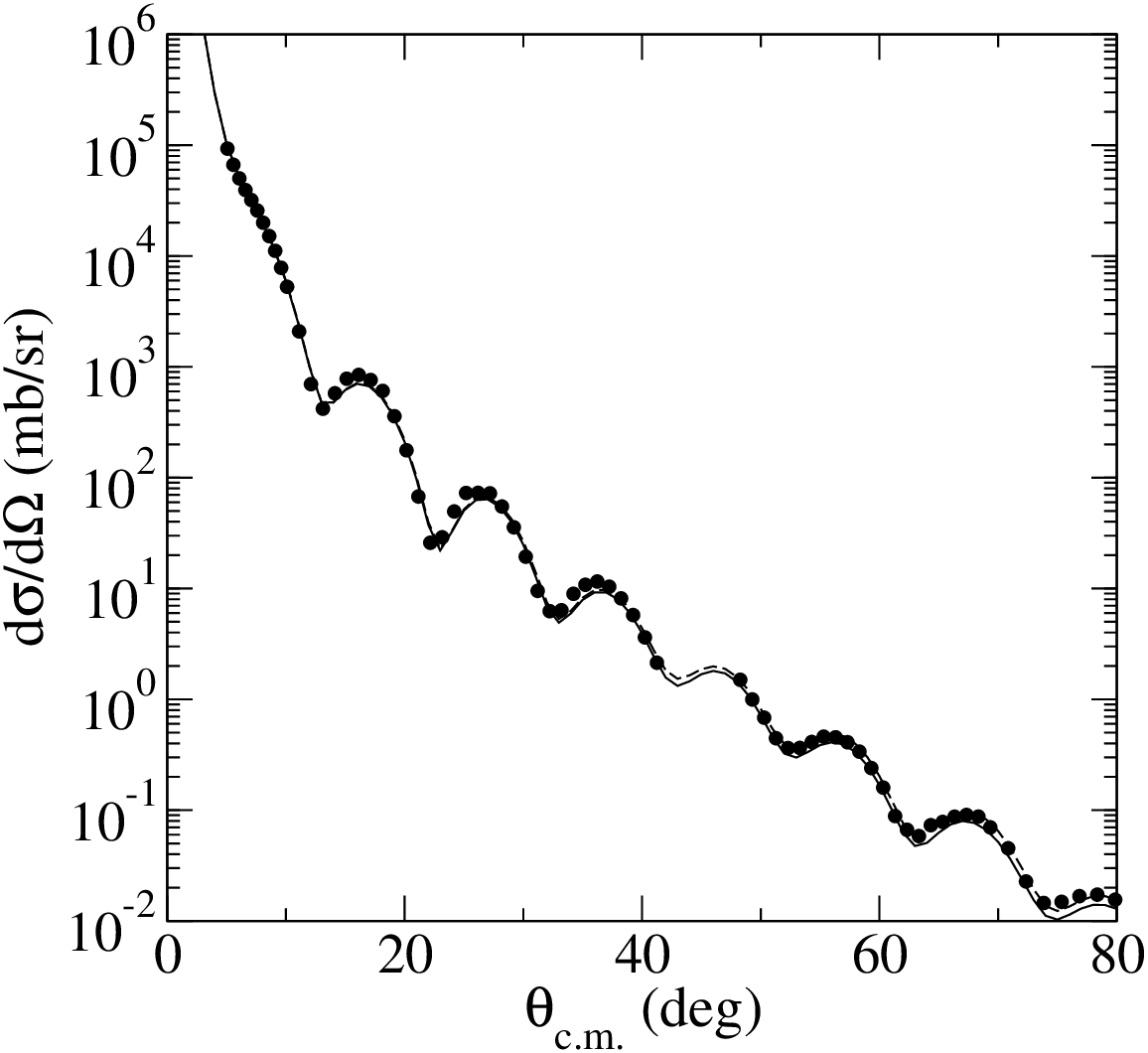}}
    \caption[]{\label{elastic}  Elastic scattering of  121~MeV protons
    from $^{208}$Pb.  The data of Nadasen  \textit{et al.} \cite{Na81}
    are  compared   to  the  results  of   $g$-folding  optical  model
    calculations made  using SkM* (solid  line) and RPA  (dashed line)
    densities.}
  \end{figure}

  As there  are no available  elastic scattering data at  135~MeV, the
  energy at which relevant inelastic scattering data exist, we compare
  results  of  the  $g$-folding  optical model  calculations  for  the
  scattering  using the  RPA and  SkM*  densities with  data taken  at
  121~MeV    \cite{Na81}.    Those    comparisons   are    shown    in
  Fig.~\ref{elastic}. Note that  these calculations are predictive: no
  adjustments at all have been made to find a better fit to the data.
  Clearly,  there is excellent  agreement between  the results  of the
  calculations  and the  data indicating  that the  densities obtained
  from both the RPA and SkM* models are reliable. The cross section is
  insensitive  to  the  difference  in  the models  which  is  largely
  confined  only to  the  surface.  This is  expected  as the  optical
  potential is dependent on the  volume integral of the density and so
  surface  effects are  relatively  minor. Hence,  to investigate  the
  surface we  turn to inelastic  scattering. However, we only  use the
  RPA  in  those analyses  as  the  Skyrme-Hartree-Fock models  cannot
  specify transitions.

  As a  first test of the  transition densities obtained  from the RPA
  for the transitions to the $2^+_1$ (4.08~MeV) and $3^-_1$ (2.61~MeV)
  states in  $^{208}$Pb, we calculate  the B($E2$) and  B($E3$) values
  for  those  transitions. The  values  obtained  from  our model  are
  0.296~$e^2$fm$^4$ and 0.692~$e^2$fm$^6$  for the B($E2$) and B($E3$)
  respectively.   Those   values    obtained   from   experiment   are
  $0.318(13)$~$e^2$fm$^4$   \cite{We84}   and  $0.611(12)$~$e^2$fm$^6$
  \cite{Go80} for the B($E2$) and B($E3$), respectively. The excellent
  level of agreement between our  model results and the data indicates
  that  no  effective charges  are  needed  in  the transition  matrix
  elements  to  correct  for  any  truncations  in  the  model  space.
  Therefore, we  may use the bare transitions  density matrix elements
  in  the calculations  of  inelastic scattering  without  the use  of
  effective charge corrections.

\begin{figure}
    \scalebox{0.4}{\includegraphics*{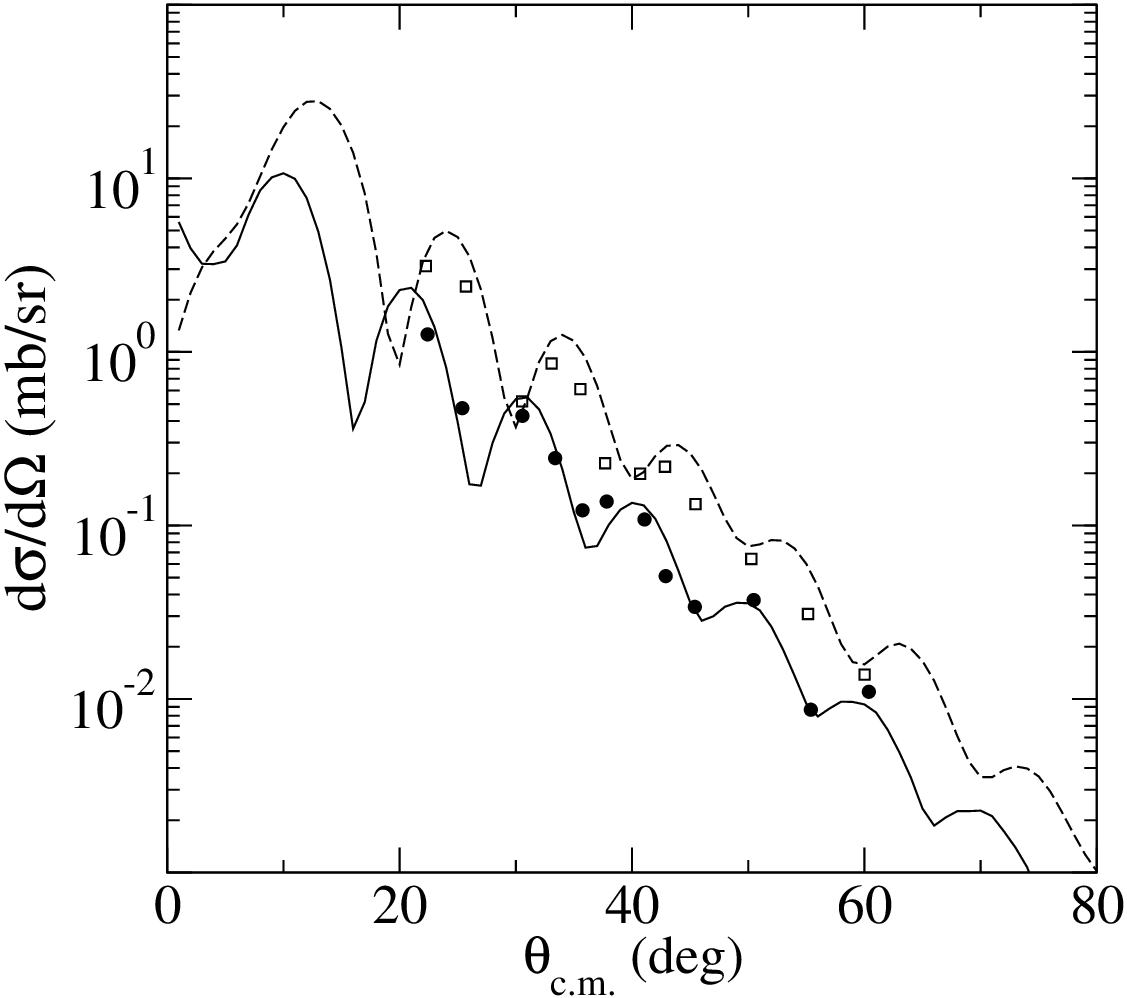}}
    \caption[]{\label{inelastic}  Inelastic scattering to  the $2^+_1$
  (4.08~MeV) and $3^-_1$ (2.61~MeV)  states in $^{208}$Pb. The data of
  Adams  \textit{et al.}  for scattering  to the  $2^+_1$  and $3^-_1$
  states are  displayed by the circles and  squares, respectively. The
  results  of the  $g$-folding calculations  made to  the  $2^+_1$ and
  $3^-_1$  states  are  displayed  by  the  solid  and  dashed  lines,
  respectively.}
  \end{figure}

  The Melbourne  $g$ matrix at  135~MeV has been used  successfully in
  analyses of scattering from $^3$He and $^{12}$C \cite{Am00}. We have
  used  that $g$  matrix in  the  calculations of  the DWA  transition
  amplitudes  for  inelastic scattering  to  the  $2^+_1$ and  $3^-_1$
  states  in  $^{208}$Pb,  for  which the  transition  densities  were
  obtained  from  our  RPA  calculation.  In  Fig.~\ref{inelastic}  we
  present results for  the scattering to both the  $2^+_1$ and $3^-_1$
  states.
  The data of Adams \textit{et  al.} to the $2^+_1$ and $3^-_1$ states
  are  denoted by  the  circles and  squares  respectively, while  the
  results of the $g$-folding calculations  made are shown by the solid
  and dashed lines, respectively.

  As with the elastic scattering, the agreement between the results of
  the  $g$-folding calculation  and  the data  for  scattering to  the
  $2^+_1$  state shown  in Fig.~\ref{inelastic}  is  excellent. Again,
  note   that   our  calculations   are   predictions.  In   contrast,
  phenomenological  analyses  of  those  data \cite{Ad80}  required  a
  deformed spin-orbit  term reflecting  the density of  $^{208}$Pb and
  for which the deformation parameters for each transition were fitted
  to each  set of data. Our  result is in general  agreement with that
  phenomenological one  without the need of  any additional spin-orbit
  components to the underlying $g$ matrix.

  In    the    case   of    scattering    to    the   $3^-_1$    state
  (Fig.~\ref{inelastic}),   the    agreement   between   our   results
  (predictions once more)  and the data is not quite  as good as those
  for the  elastic scattering or  the $2^+_1$ transition.  The earlier
  phenomenological  calculation \cite{Ad80} does  better by  virtue of
  fitting the deformation parameters  of that optical potential to the
  data  being described.  Note  that  both the  data  and our  results
  naturally follow the phase rule of Blair \cite{Am65}.

  It  is   insightful  to  compare   our  results  with  those   of  a
  semi-microscopic calculation \cite{Pe80} in  which a $t\rho$ form of
  the  optical  potential  with  transition  densities  obtained  from
  analyses of inelastic electron scattering was used. The $t$ matrices
  used  in   that  analyses  were   those  of  Love   \textit{et  al.}
  \cite{Lo78}. Those  analyses gave quite  a good reproduction  of the
  $3^-_1$ transition  when the requisite densities  were obtained from
  electron  scattering data, and  when just  the central  and two-body
  spin-orbit components  of the $t$  matrix were chosen. They  do not,
  however, reproduce the $2^+_1$ transition. Agreement with those data
  is achieved only when the central term of the $t$ matrix is used. In
  part,  that may be  due to  a problem  in extracting  the transition
  density  from the  available electron  scattering  data \cite{Pe80}.
  However,  the problem  may  also lay  in  the isoscalar  assumption,
  equating  the  proton  and  neutron  densities, which  was  used  in
  constructing these optical potentials.  The reason as given was that
  the  isoscalar parts  of the  interaction are  much larger  than the
  isovector  parts.  Yet that  assumption  fails  to conserve  neutron
  number  and so  the density  dependence of  their potentials  is not
  correct. This  would be especially problematic  in proton scattering
  at these energies. As proton scattering probes primarily the neutron
  density it is  important to ensure that the  correct neutron density
  is used. That is the case with the present calculations.

  We have  predicted, using a fully  microscopic parameter-free model,
  the elastic and inelastic  scattering of intermediate energy protons
  from $^{208}$Pb.  The densities used for the  elastic scattering and
  transitions  were obtained  from an  RPA calculation  using  the D1S
  effective $NN$  interaction. The RPA model allowed  for an effective
  no-core microscopic model description of the spectrum of $^{208}$Pb.
  The skin thickness obtained using the RPA wave functions is 0.13~fm,
  as  compared to  0.17~fm obtained  from  the SkM*  model. Those  RPA
  densities were  folded with the  Melbourne $g$ matrices to  give the
  microscopic  optical potentials  needed to  describe  the scattering
  without  any  fitting of  parameters  to  the  data being  described
  \textit{a  posteriori}. Excellent  agreement has  been  obtained for
  both elastic and inelastic scattering commensurate with descriptions
  of elastic  and inelastic scattering  for much lighter nuclei.  As a
  result, we are  pursuing analyses of inelastic scattering  up to and
  including  excitation of  the  $12^+$ state.  The present  analyses,
  together  with  those   obtained  previously,  constraing  the  skin
  thickness to  lie in  the range  $0.13 < S  < 0.17$~fm.  While these
  results were obtained using the pure RPA, they give encouragement to
  studying the structures of heavy nuclei with nucleon scattering when
  one generalises also to the  use of the QRPA or Generator Coordinate
  Method \cite{Ri80}. The information  gained on the neutron densities
  of such  nuclei may  then be used  to study effectively  the neutron
  equation of state.

  \bibliography{pb_scatt_fin}

\end{document}